\documentclass[12pt]{iopart} 

\usepackage{graphicx} 
\usepackage{bm}

\begin{document} 

\title{Hot-electron noise suppression in n-Si via the Hall effect} 

\author{Francesco Ciccarello} 

\address{CNISM and Dipartimento di Fisica e 
Tecnologie Relative dell'Universit\`{a} degli Studi di Palermo, 
Viale delle Scienze, Edificio 18, I-90128 Palermo, Italy} 
\ead{ciccarello@difter.unipa.it} 

\author{Salvatore Zammito} 

\address{CNISM and Dipartimento di Fisica e 
Tecnologie Relative dell'Universit\`{a} degli Studi di Palermo, 
Viale delle Scienze, Edificio 18, I-90128 Palermo, Italy} 

\author{Michelangelo Zarcone} 

\address{CNISM and Dipartimento di Fisica e 
Tecnologie Relative dell'Universit\`{a} degli Studi di Palermo, 
Viale delle Scienze, Edificio 18, I-90128 Palermo, Italy} 

\begin{abstract} 
We investigate how hot-electron fluctuations in n-type Si are 
affected by the presence of an intense (static) magnetic field in a 
Hall geometry. By using the Monte Carlo method, we find that the 
known Hall-effect-induced redistribution of electrons among valleys 
can suppress electron fluctuations with a simultaneous enhancement 
of the drift velocity. 
\end{abstract} 

\pacs{} 

\section{Introduction} 

The study of hot-electron transport properties of semiconductor 
materials has witnessed impressive progresses in the last decades. 
Clearly, these have been motivated by the well-known technologic 
relevance of semiconductors, together with the unceasing 
miniaturization process of electronic devices. In semiconductor 
micro-devices electrons may be accelerated by electric fields of the 
order of even several kV/cm, so the transport is in general highly 
non-ohmic and deviations from thermal equilibrium are significant. 
Mainly due to the formidable mathematical difficulties in solving 
the Boltzmann equation in such cases, the Monte Carlo (MC) method 
for the direct simulation of charge carriers' motion became what is 
probably the most popular numerical tool for approaching problems of 
this sort \cite{MC}. While most of the studies in this area focused 
on transport in the presence of electric fields, some MC 
investigations addressed the case where both an electric and a 
magnetic field are present in compounds such as GaAs 
\cite{boardman,djikstra,ciccarello}, InSb 
\cite{lituani,brazis,warmenbol, warmenbol2}, Si \cite{raguotis} and 
GaN \cite{albrecht}. In particular, two of us have recently 
investigated hot-electron noise in n-GaAs in the Hall geometry for 
electric fields of the order of some kV/cm and magnetic-field 
strengths up to 2 T \cite{ciccarello}. The presence of the magnetic 
field was shown to affect the electron-velocity autocorrelation 
function and the noise spectrum of velocity fluctuations in a 
non-trivial way. In particular, the magnetic field was shown to give 
rise to an attenuation of electron fluctuations around the drift 
velocity for electric-field strengths intense enough to yield a 
significant population of $L$-valleys \cite{ciccarello}. 

In a multivalley semiconductor, a significant contribution to 
nonlinearity comes from the carriers' distribution among valleys 
with different effective masses. This is also true when all the 
valleys are energetically equivalent but, due to conduction-band 
anisotropy such as in Si, they have different mass tensor. The 
electron heating by the applied electric field is higher in valleys 
with lower mass along the field direction and thus such hot valleys 
are less populated. However, if a magnetic field is added to the 
applied electric field according to a Hall geometry, the above 
behaviour can be reversed: the hot (cold) valleys in the 
zero-magnetic-field case now become the cold (hot) ones. Such effect 
was demonstrated long ago in n-Si at T=77 K, both experimentally 
and theoretically, and shown to be able to 
enhance conductivity \cite{sarbey}. To the best of our knowledge, 
however, no analysis of how electron velocity fluctuations are 
affected was carried out so far. Regardless, the problem is 
physically intriguing. Indeed, it is well-known that the application 
of a static electric field along crystallographic directions such as 
$\langle 100 \rangle$ (or equivalent ones) in n-Si give rise to the 
so called \emph{partition noise}: Electrons in valleys with 
different effective masses along the field direction have also 
different average  velocities. As the drift velocity has of 
course an intermediate value, this implies an extra amount of 
velocity fluctuations. The typical signature of such phenomenon is 
the presence of a long-time tale in the longitudinal autocorrelation 
function of velocity fluctuations (whose length is dictated by the 
$f$-type intervalley scattering rate) 
\cite{brunetti-PhyRev,brunetti-JAP,fauquembuergue}. The main 
motivation behind this work is to establish whether in a Hall 
geometry the aforementioned Hall-effect-induced redistribution of 
electrons among different valleys is able to attenuate 
partition noise.

When, in addition to the electric field, a perpendicular magnetic 
field is also applied, two main effects occur. First, the 
free-flight electron motion in the momentum space now takes place 
along closed orbits and thus the amount of gainable energy in a 
given free-flight is upper-bounded \cite{boardman,ciccarello}. 
Second, the Hall field adds to the applied electric field and at 
strong enough magnetic fields the former may even exceeds the 
latter. In a multivalley semiconductor, both the magnetic and the 
Hall fields can in general be sensed differently by electrons moving 
in different valleys and thus the conduction properties in the 
hot-electron regime may be significantly affected \cite{sarbey}. 
Here, we shall show that the drift-velocity enhancement shown in 
Ref.~\cite{sarbey} may occur with a simultaneous decrease of 
velocity fluctuations and partition noise. 

This paper is organized as follows. In Sec.~\ref{system_approach}, we 
describe the model and parameters used for the conduction band of 
n-Si and the various scattering processes as well as the approach we 
followed to tackle the problem. Sec.~\ref{par_centrale} is devoted 
to the central result of this work, i.e. the reduction of partition noise and of the 
relative importance of velocity fluctuations 
compared to the drift velocity. Additional plots illustrating the 
mechanism behind are presented and discussed. Finally, in Sec.~\ref{concl} we draw our 
conclusions and outline the open problems. 

\section{Model and approach} \label{system_approach} 

\begin{figure} 
\centering 
\includegraphics [scale=0.3]{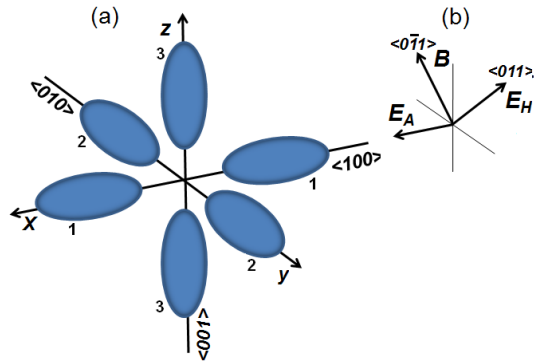} \caption{(a) Sketch of the 
valleys 1, 2 and 3 
along the crystallographic directions $\langle 100\rangle$, $\langle 
010\rangle$ and $\langle 001\rangle$, respectively. (b) Hall 
geometry with $\mathbf{E}_A$, $\mathbf{E}_H$ and $\mathbf{B}$ along 
the crystallographic directions $\langle 100\rangle$, $\langle 
011\rangle$ and $\langle 0\bar{1}1\rangle$, respectively.} 
\label{Fig1} 
\end{figure}

We have modeled the conduction band of Si as three pairs of 
nonparabolic ellipsoidal valleys along the $\langle 100 \rangle$, 
$\langle 010 \rangle$ and $\langle 001 \rangle$ directions, to be 
referred to as valleys 1, 2 and 3, respectively [see Fig.~\ref{Fig1}(a)]. The 
$\langle 100 \rangle$, $\langle 010 \rangle$ and $\langle 001 
\rangle$ directions are taken as the $x$, $y$ and $z$-axis, 
respectively, of the reference frame here adopted [see Fig.~\ref{Fig1}(a)]. 
For each valley, we take as longitudinal and transverse effective 
masses $m_l=0.916\,m_0$ and $m_t=0.19\,m_0$, respectively, and 
nonparabolicity factor $\alpha=0.5$ eV$^{-1}$. We have included six 
intervalley-phonon scattering processes, three of $f$-type and three 
of $g$-type, as well as acoustic intravalley scattering 
\cite{jacoboni-param} with the associated phonon equivalent 
temperatures and potentials used in Ref.~\cite{brunetti-JAP}. 
Assuming a pure enough sample, no impurity scattering has been 
included \cite{brunetti-JAP}. 

The MC procedure for tackling hot-electron transport in crossed 
electric and magnetic fields in a semiconductor bulk was first 
suggested by Boardman \emph{et al.} \cite{boardman} and used in 
Ref.~\cite{ciccarello} for the case of GaAs. The basic idea is to 
derive the Newtonian time-evolution of the electron wave vector 
$\mathbf{k}(t)$ during free flights in the presence of the static 
magnetic field $\mathbf{B}$ and the \emph{total} electric field 
$\mathbf{E}$. Unlike in the absence of magnetic field, the latter is 
the superposition of the applied electric field and the Hall field 
according to $\mathbf{E}=\mathbf{E}_A+\mathbf{E}_H$. To overcome the 
difficulty that for set values of $\mathbf{E}_A$ and $\mathbf{B}$ 
the Hall field $\mathbf{E}_H$ is initially unknown, the total 
electric field $\mathbf{E}$ is used as an independent parameter, 
together with $\mathbf{B}$, and the simulation is run in order to 
compute the drift velocity $\mathbf{v}_d$. The component of the 
total electric field $\mathbf{E}$ along (orthogonal to) the 
direction of $\mathbf{v}_d$ yields $\mathbf{E}_A$ ($\mathbf{E}_H$) 
\cite{boardman}. This simple approach is fruitful in the case of 
isotropic materials (such a GaAs), when the transport properties 
depend solely on the magnitude of $\mathbf{E}$. However, due to 
electron-mass anisotropy, the procedure cannot be adopted in 
materials such as Si where $\mathbf{v}_d$ depends on the direction 
of the electric field as well \cite{raguotis}. In our analysis we 
thus treated both $E_A$ and $E_H$, i.e. the two components of 
$\mathbf{E}$ in the plane orthogonal to $\mathbf{B}$, as independent 
parameters of the MC simulation. For given $E_A$ and $B$, different 
simulations were run, each with a different value of $E_H$. The 
correct Hall field is the one yielding $\mathbf{v}_d$ along 
$\mathbf{E}_A$ \cite{raguotis}. Regarding the Hall geometry, in this 
work we address the case where $\mathbf{E}_A$, $\mathbf{E}_H$ and 
$\mathbf{B}$ lie along the crystallographic directions $\langle 
100\rangle$, $\langle 011\rangle$ and $\langle 0\bar{1}1\rangle$, 
respectively [see Fig.~\ref{Fig1}(b)]. Such geometry has the advantage that 
the fields are symmetric with respect to valleys 2 and 3 and thus 
such groups of valleys necessarily exhibit the same behaviour. This 
will simplify our discussion. 

The time-evolution of the electron wave vector in a given valley 
during a free flight is governed by the Newton's law 
\begin{equation}\label{NL} 
\hbar \dot{\mathbf{k}}=e\left[\mathbf{E}+\mathbf{v(\\ 
k)}\times \mathbf{B} \right], 
\end{equation} 
where $\mathbf{v(\\ 
k)}\!=\!\frac{1}{\hbar}\bigtriangledown_{\mathbf{k}}\varepsilon(\mathbf{k})$ 
is the electron group velocity and the energy-wave vector 
relationship that accounts for nonparabolicity effects is 
\cite{conwell} $\varepsilon(1+\alpha \varepsilon)=\hbar^2/2\, 
(k_l^2/m_l+k_t^2/m_t)$ [$k_l$ ($k_t$) are the longitudinal 
(transverse) components of the wave vector]. In order to fully 
account for the nonparabolicity effects which are rather important 
in Si \cite{jacoboni-param}, Eq.~(\ref{NL}) has been solved 
numerically at each free flight. Despite Eq.~(\ref{NL}) has no 
trivial analytic solution, some qualitative predictions can be 
established. Indeed, it is clear that, similarly to the isotropic 
case \cite{boardman}, the free-flight electron motion in the 
momentum space takes place along closed orbits around a displaced 
center. In the isotropic case the period of motion is dictated by 
the cyclotron frequency $\omega_c=e B/m^*$ \cite{boardman, ciccarello}. Here, 
due to band anisotropy which implies the presence of the two 
effective masses $m_l$ and $m_t$, two cyclotron frequencies are 
exhibited. In addition, unlike the circular nature of motion in the 
isotropic case, in the present case the orbits are in general 
elliptical. 

As the fields are static and we are concerned with steady-state 
transport in bulk n-Si, the results presented in this work were 
obtained by simulating single-particle time-histories for times up 
to $\mu$s. The treatment of the scattering processes was performed 
in the usual way \cite{MC}. In this paper, we consider electric 
fields of the order of some kV/cm and magnetic-field strengths of 
the order of some T. We have checked that the product between the 
cyclotron frequencies and the typical electron relaxation time 
exhibits moderate values, which justifies that the magnetic-field 
strengths here considered do not give rise to quantum effects as we 
have assumed. 

Finally, both the drift velocity $v_d$ and the variance of velocity 
fluctuations $\langle \delta v^2\rangle$ were computed from a 
sampling of the electron velocity along the applied-electric-field direction. The results presented in 
this paper were obtained 
by setting the temperature to 77 K. 

\section{Noise suppression} \label{par_centrale} 

\begin{figure}[htbp] 
\begin{center} 
{\includegraphics[scale=0.30]{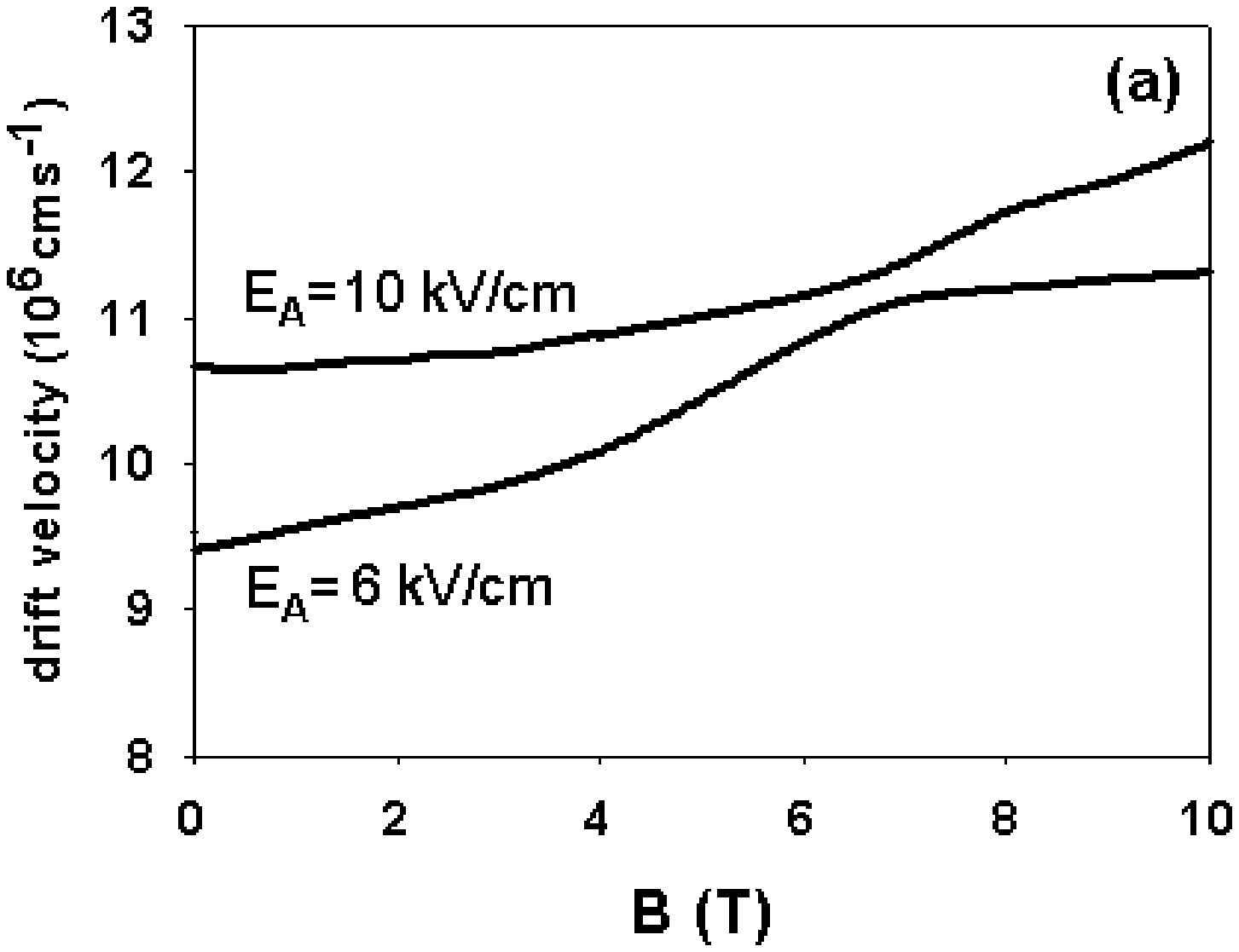}} 
{\includegraphics[scale=0.30]{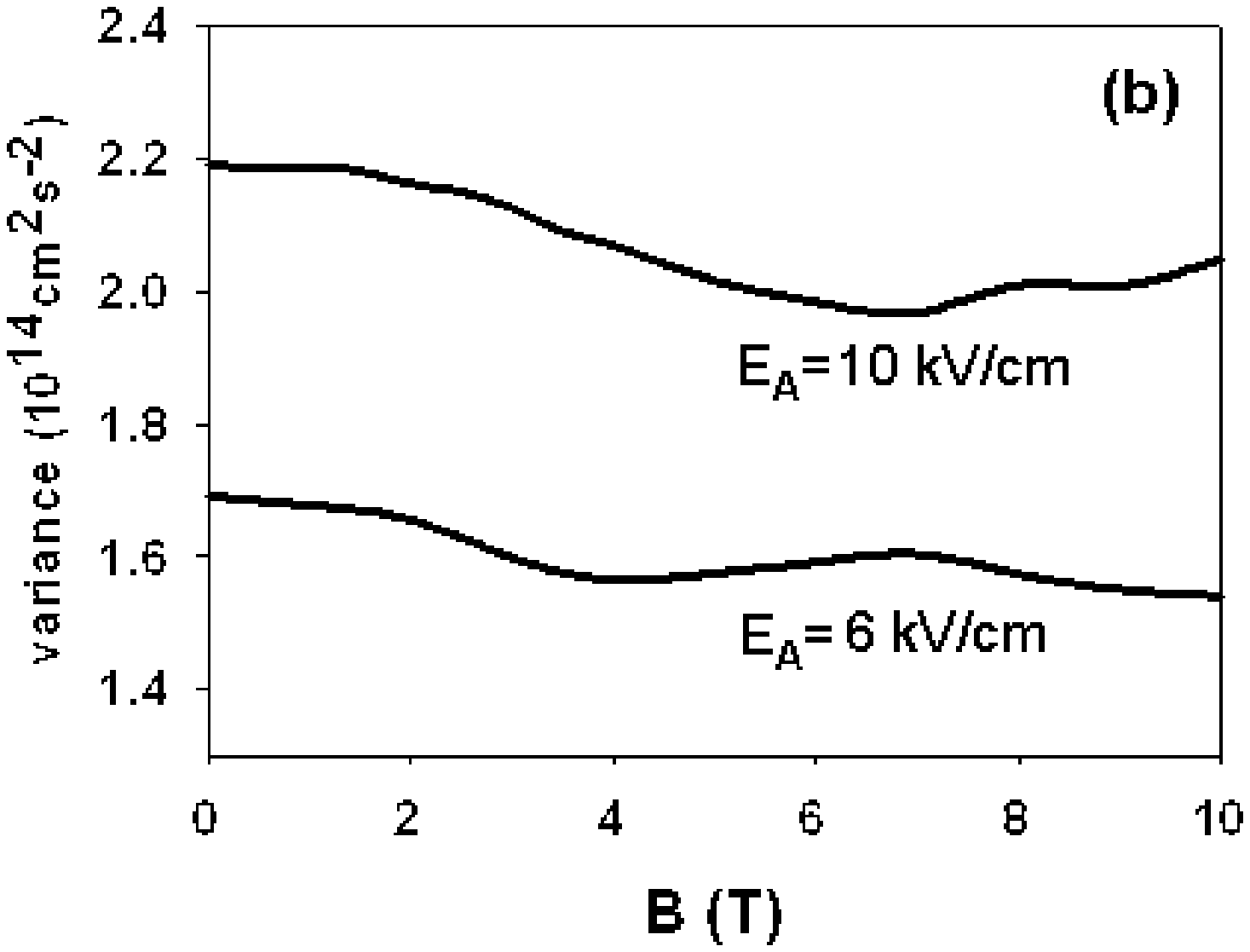}}{\includegraphics[scale=0.30]{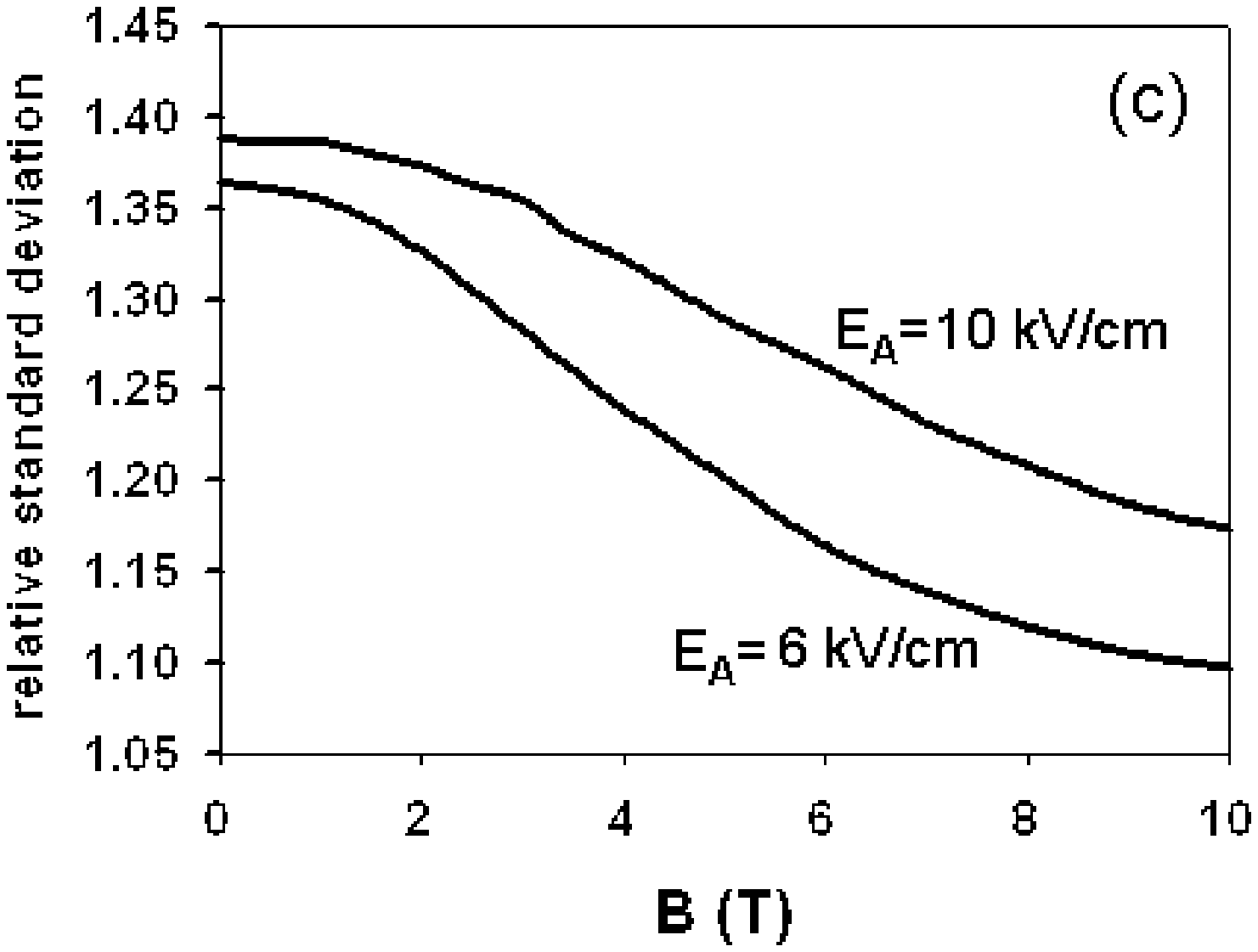}} 
\caption{\footnotesize {(a) Drift velocity $v_d$ (10$^6$ cm/s) vs. magnetic field $B$. 
(b) Velocity variance $\langle \delta v^2\rangle$ (10$^{14}$ cm/s) vs. $B$. (c) Relative 
standard deviation $\sqrt{\langle \delta v^2\rangle}/v_d$ vs. B. The 
set values of the applied electric field are $E_A$=6, 10 kV/cm.}}\label{Fig2} 
\end{center} 

\end{figure}

\begin{figure}[htbp] 
\begin{center} 
{\includegraphics[scale=0.30]{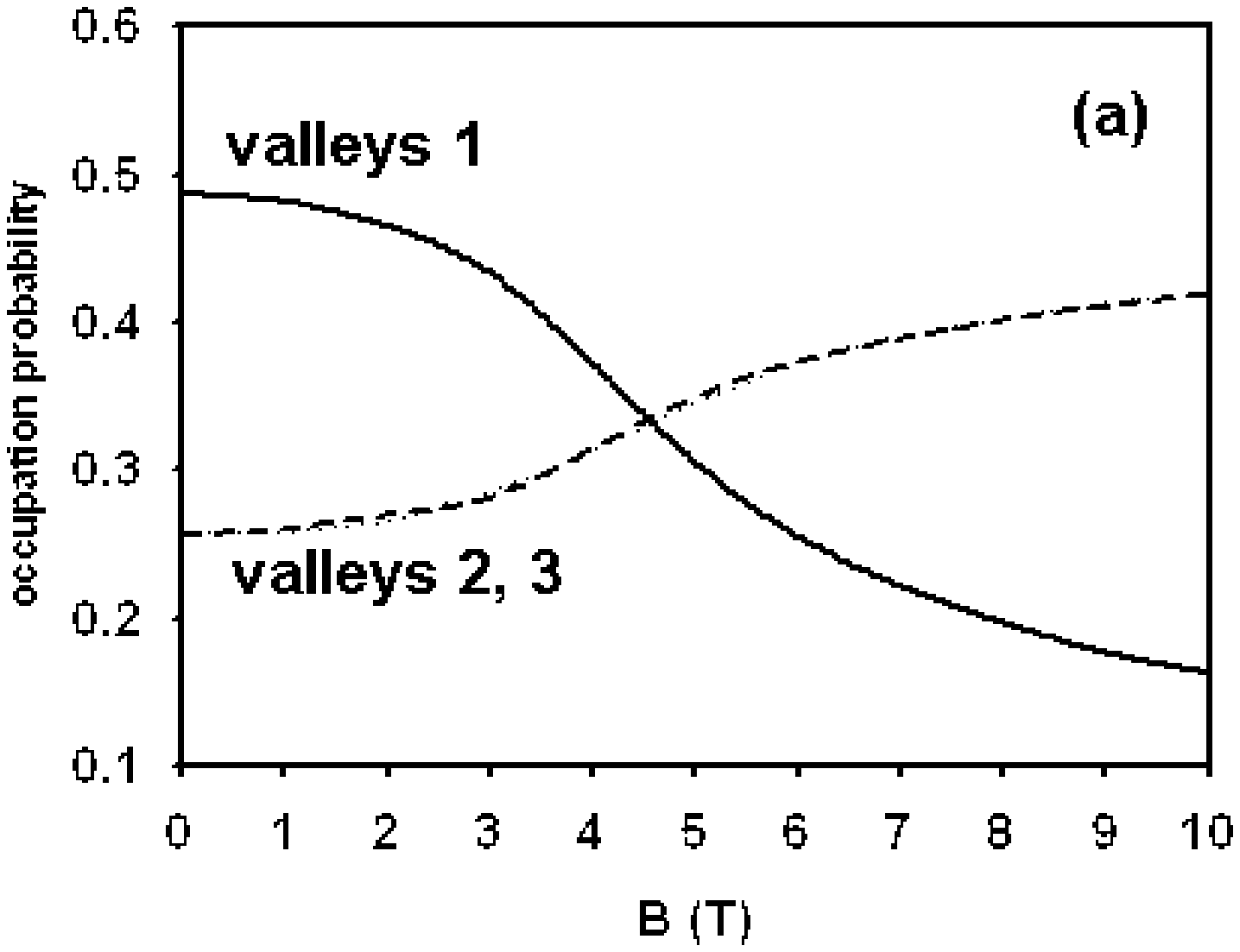}} 
{\includegraphics[scale=0.30]{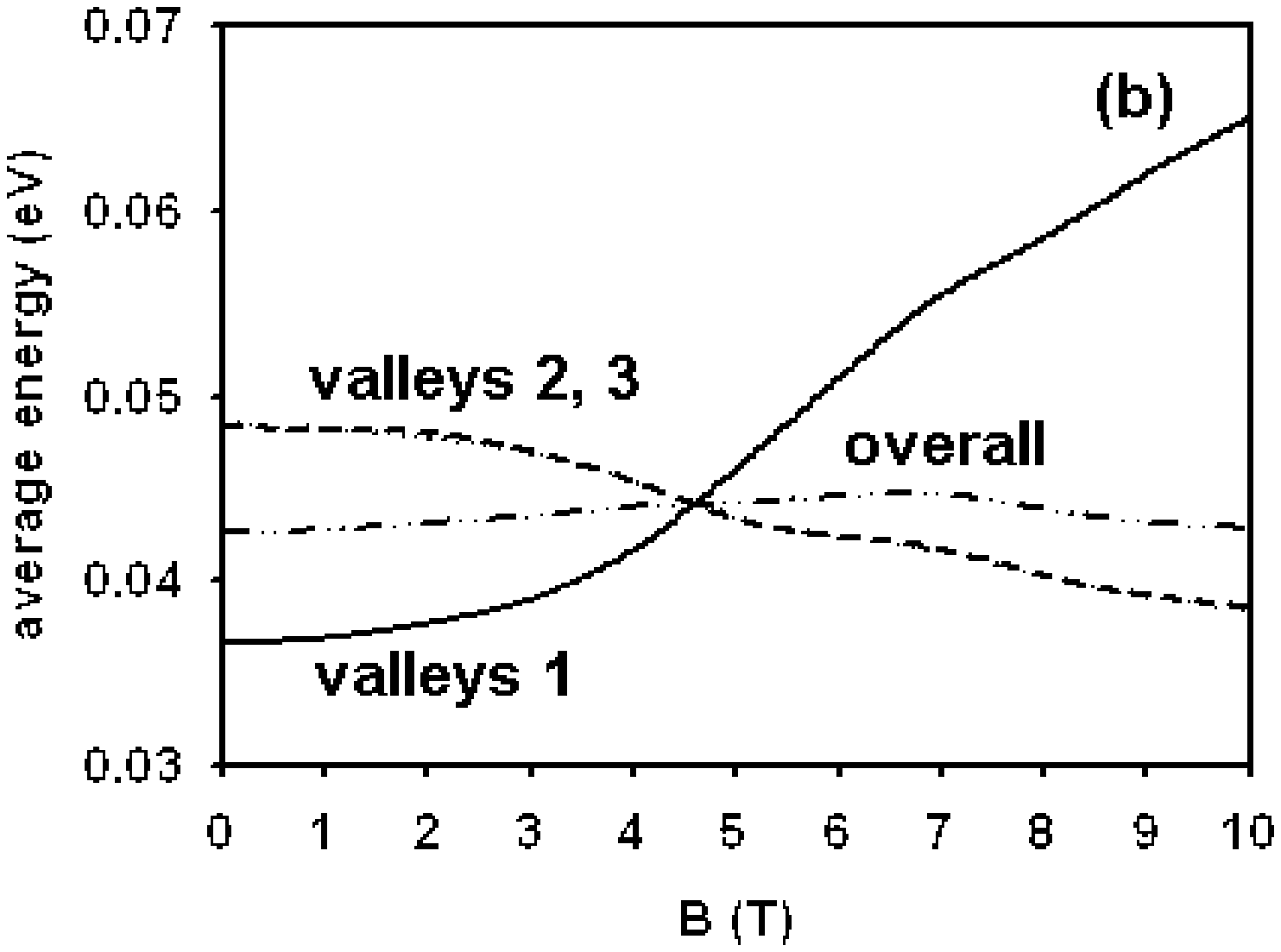}}{\includegraphics[scale=0.30]{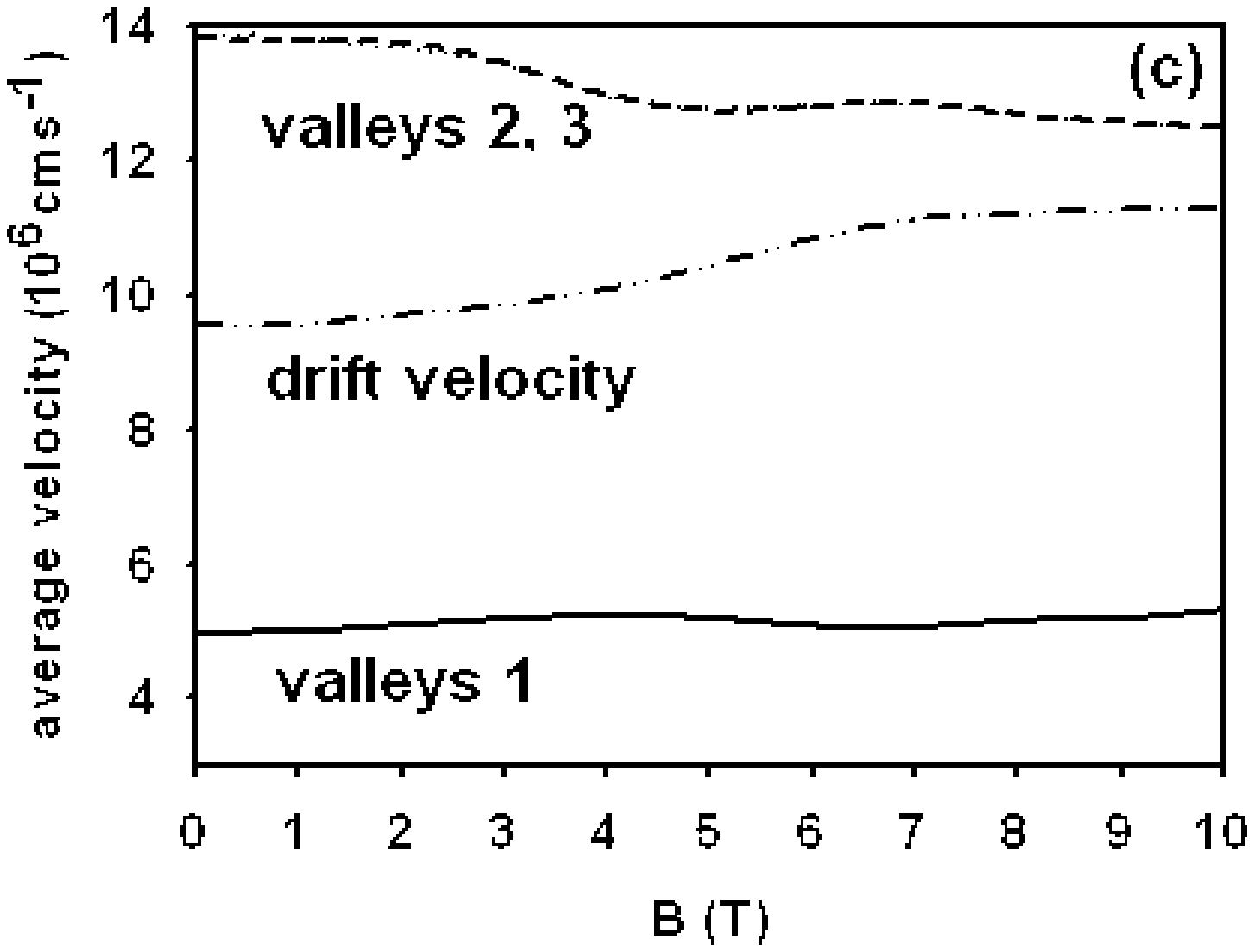}} 
\caption{\footnotesize {(a) Valley occupation probabilities vs. magnetic field $B$. 
(b) Valley average energy and overall average energy vs. $B$. (c) Valley average velocities along 
the applied electric field and drift 
velocity vs. $B$. We have set the applied-electric-field strength 
to $E_A$=6 kV/cm.}} 
\end{center}\label{Fig3} 
\end{figure}

In Fig.~\ref{Fig2} we plot the drift velocity $v_d$ (a), the variance 
$\langle \delta v^2\rangle$ (b) and the relative standard deviation 
$\sqrt{\langle \delta v^2\rangle}/v_d$ (c), respectively, against 
the magnetic field strength for $E_A$=6, 10 kV/cm. 
Notice that, for a given applied electric field, the drift velocity 
$v_d$ grows with $B$, which confirms the magnetic-field-induced 
increase of conductivity predicted in Refs.~\cite{sarbey}. Such 
interpretation is strengthened by the behaviour of the populations 
and mean energies of the three group of valleys for increasing 
values of $B$ plotted in Figs.~\ref{Fig3}~(a) and (b), respectively, 
in the case $E_A$=6 kV/cm. 
At low magnetic fields, when the dominating effect is 
that due to the electric field, valleys 1 are the coldest, and thus 
the most populated, given that they have the largest effective mass 
along the applied electric field $m_x\!=\!m_l$ [see Figs.~\ref{Fig1}~(a) and (b)]. 
As $B$ grows, however, valleys 2 and 3 are progressively cooled, 
whereas valleys 1 are heated. At a threshold magnetic-field strength 
($\simeq$4.5 T for the case $E_A$=6 kV/cm considered in 
Fig.~\ref{Fig3}) all the valleys exhibit the same average energies 
and populations. Above such threshold field valleys 2 and 3 (valleys 
1) now become the coldest and most populated (the hottest and less 
populated), which further confirms the effects investigated in 
Ref.~\cite{sarbey}. Concerning the overall average 
energy, it can be noticed from Fig.~\ref{Fig3}(b) that for growing 
magnetic fields such a quantity exhibits an initial moderate rise 
(with a maximum increase lower than $\simeq$5\%) followed by a 
decrease. This trend can be understood as follows. For weak magnetic 
fields, valleys 2 and 3 are still the hottest and thus an increase 
in their population produces a higher average energy. For an intense 
enough magnetic field, as the effect of the electric field becomes 
negligible and the Lorentz-force work is null the electron energy 
reduces to its thermal value, which explains the decrease. 

Figs.~\ref{Fig2} ~(a) and (b) show that, while the drift-velocity is 
enhanced, the variance is lowered by the magnetic field compared to 
the zero-magnetic-field case, which indicates that the total noise 
power is significantly attenuated. Therefore, as a simultaneous 
conductivity enhancement takes place [cfr. Fig.~\ref{Fig2}(a)] the 
presence of the magnetic field reduces the relative importance of 
velocity fluctuations. This result is strengthened by 
Fig.~\ref{Fig2}(c) that shows an essentially monotonous decrease of 
the relative standard deviation $\sqrt{\langle \delta 
v^2\rangle}/v_d$ in the considered magnetic-field range, the 
reduction being able to be as pronounced as $\simeq$20\%. In short, 
the application of the magnetic field is able to yield both a 
more intense and cleaner signal. We believe this is a rather 
remarkable result. 

It is worth pointing out that, as expected, larger magnetic-field strengths 
are required for growing electric fields in order for such effects 
to be exhibited, which is witnessed by a comparison between the 
plots obtained with different values of $E_A$ in Fig.~\ref{Fig2}. 

In order to shed light on the mechanism behind noise suppression, in 
Fig.~\ref{Fig3}(c) we analyze the behaviour of the valley average 
velocities along the longitudinal direction, i.e. that of the 
applied electric field ($x$-axis). First, notice that such 
velocities are rather weakly affected by the magnetic field. As in 
the zero-magnetic-field case, electrons visiting valleys 2 and 3 
move faster than those in valleys 1. Nonetheless, the magnetic field 
induces a significant transfer of electrons from valleys 1 to 
valleys 2 and 3 [cfr. Fig.~\ref{Fig3}(a)], which explains why drift 
velocity is enhanced [cfr. Figs.~\ref{Fig2}(a) and \ref{Fig3}(c)]. At the same time, as 
valleys 2 and 3 are significantly occupied even in the absence of 
magnetic field, such a transfer must necessarily yield a reduced 
average discrepancy between valley velocities and (increased) drift 
velocity, i.e. suppression of partition noise takes place. It should 
be pointed out that, unlike in the absence of magnetic field, in a 
Hall geometry the electron velocity has intrinsic cyclotron 
oscillations around its average value dictated by the effective 
masses and the magnetic-field strength (see Sec. 
\ref{system_approach} and Refs.~\cite{boardman,ciccarello}). It 
follows that at intense magnetic fields a relevant contribution to 
variance comes from such oscillations and thus it might be more 
appropriate in the present case using the terminology ``variance of 
velocity \emph{deviations}". The effects of cyclotron oscillations 
therefore compete with those of partition-noise suppression and the 
velocity variance in Fig.~\ref{Fig2}(b) results from the interplay 
between the two. This is likely to be the reason behind the 
non-monotonous behaviour in Fig.~\ref{Fig2}(b). Anyway, it appears 
reasonable concluding that mere velocity \emph{fluctuations} are 
actually attenuated. 

\begin{figure} 
\centering 
\includegraphics [scale=0.4]{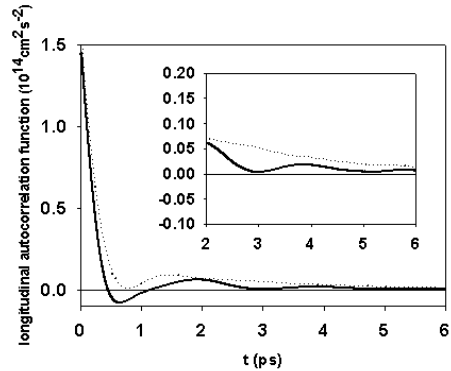} 
\caption{Longitudinal autocorrelation function with $B$=4.5 T (solid line) and 
in the absence of magnetic field (dotted line). In the insert, a zoom of the 
time-tale is shown.} \label{Fig4} 
\end{figure}

In order to provide further evidence of partition-noise suppression, 
in Fig.~\ref{Fig4} we set $E_A$=5 kV/cm and compare the longitudinal autocorrelation 
function in the absence of magnetic field (dotted line) with that obtained with 
a magnetic field $B$=4.5 T 
(solid line). 
As expected \cite{ciccarello}, in the Hall geometry damped oscillations 
are exhibited as a signature of cyclotron motion. Nonetheless, the oscillations' 
center is only slightly above zero. Indeed, a comparison of the time tails with and 
without magnetic field [see the insert in Fig.~\ref{Fig4}] shows a magnetic-field-induced 
attenuation of the long-time-tail effects (remind that these are a well-known 
signature of partition noise \cite{brunetti-PhyRev,brunetti-JAP,fauquembuergue}). 

\section{Conclusions} \label{concl} 

In summary, in this paper we have investigated how hot-electron noise in n-Si 
is affected by the presence of a magnetic field in a Hall geometry. 
In agreement with previous works, we have found that the presence of the 
magnetic field gives rise to a redistribution of electrons among valleys 
and enhancement of conductivity. Remarkably, we have found that 
the drift-velocity enhancement for a given applied electric field occurs 
with a simultaneous decrease of variance of velocity fluctuations, 
i.e. the total noise power, and thus of the relative importance of velocity fluctuations. 
We have shown that such effect results from the magnetic-field-induced transfer 
of electrons from heavy to light valleys (with respect to the direction of the 
applied electric field), which causes a significant attenuation of 
partition noise. We have provided further evidence of partition-noise suppression 
by comparing the longitudinal autocorrelation functions obtained 
with and without magnetic field. We have found that the well-known 
long-time tail effects due to partition-noise are significantly 
attenuated in the Hall geometry. 

In short, it seems reasonable summarizing the results obtained 
in this work by stating that the magnetic field is able to introduce \emph{order} 
in the hot-electron dynamics. This circumstance results in a 
conductive state with higher performances compared to the case where 
only the electric field is present. We believe the phenomena presented in this work 
may be attractive in order to design novel strategies for minimizing noise in 
semiconductor materials exposed to intense electric fields. 

Some open questions still remain, such as: What is the optimal Hall geometry 
for minimizing noise and/or maximizing conductivity? Indeed, it is well-known 
that in Si, and in general in anisotropic semiconductors, different 
transport features are obtained with different electric-field geometries. 
Evidently, when a magnetic field is simultaneously applied, the number 
of non-trivial Hall geometries giving rise to different 
conductive regimes is expected to be larger. Indeed, a part of our ongoing 
work is aimed at establishing the 
importance of geometry in these phenomena. 

Finally, as already mentioned, when quantities such as the variance 
of velocity fluctuations or the velocity autocorrelation function are computed 
in a Hall geometry, the outcomes are affected by both mere fluctuations 
and intrinsic cyclotron oscillations. Therefore, if one is concerned with 
a mere noise analysis, it would be desirable to filter out somehow the effect 
of cyclotron motion. The derivation of an appropriate statistical procedure 
for accomplishing such a task is under ongoing investigations. 

\section{Acknowledgements} 

The authors are indebted to M G Santangelo (ETH, Z\"{u}rich) for the 
help offered in the search of old papers. This work makes use of results produced by the PI2S2 
Project managed by the 
Consorzio COMETA, a project cofunded by the Italian Ministry of University 
and Research (MIUR) within the Piano Operativo Nazionale "Ricerca 
Scientifica, Sviluppo Tecnologico, Alta Formazione" (PON 2000-2006).

\end{document}